\documentclass[twocolumn,prl]{revtex4}
\usepackage{amssymb}
\usepackage{amsfonts}
\usepackage{amsmath}
\usepackage{graphicx}
\usepackage{graphics}
\usepackage{bm}
\usepackage{mathdots}
\usepackage{calc}
\usepackage{dcolumn}
\newcommand{\mE}{{\mathcal E}}

\newcommand{\tr}{{\mbox{Tr}}}
\usepackage{pstricks}
\usepackage{pst-plot}
\begin{document}
\title{Optimal joint measurement of  two observables of a qubit
}

\author{Sixia Yu and C.H. Oh}
%\affiliation{Hefei National Laboratory for Physical Sciences at Microscale and Department of Modern Physics, University of Science and Technology of China, Hefei, Anhui 230026, China}
\affiliation{Centre for Quantum Technologies, National University of Singapore, 2 Science Drive 3, Singapore 117542}
%\author{C.H. Oh}
%\affiliation{Centre for Quantum Technologies, National University of Singapore, 2 Science Drive 3, Singapore 117542}
%\affiliation{Physics Department, National University of Singapore, 2 Science Drive 3, Singapore 117542}

\begin{abstract}
Heisenberg's uncertainty relations for measurement quantify how well we can jointly measure two complementary observables and have attracted much experimental and theoretical attention recently. Here we provide an exact tradeoff between the worst-case errors in measuring jointly two observables of a qubit, i.e., all the allowed and forbidden pairs of errors, especially asymmetric ones, are exactly pinpointed. For each pair of optimal errors we provide an optimal joint measurement  that is realizable without introducing any ancilla and entanglement. Possible experimental implementations are discussed and Toronto experiment [Rozema {\it et al.}, Phys. Rev. Lett. 109, 100404 (2012)] can be readily adapted to an optimal joint measurement of two orthogonal observables. 
\end{abstract}

\maketitle

{\it Introduction.--- } 
One distinguishing feature of quantum theory is the existence of observables that cannot be jointly measured in a single measuring apparatus. These incompatible observables, also referred to as  noncommuting or complementary observables, describe mutually exclusive properties of the quantum system  in Bohr's complementarity \cite{bohr} and define different measurement contexts for quantum contextuality \cite{ks}. However, incompatible observables can be approximately measured in a single apparatus by allowing some errors. Of course there are some tradeoffs among the allowed errors, e.g., the errors cannot be both zero for two incompatible observables. And these tradeoffs are referred to as Heisenberg's uncertainty relation \cite{heis} for measurement. In comparison Heisenberg's uncertainty relations for preparation \cite{Kennard1927, Robertson1929, Yu2013} do not involve any efforts of measuring jointly two or more incompatible observables.

Ozawa \cite{Ozawa2003} established the first example of universally valid uncertainty relation for measurement, using the root-mean-square errors to characterize the error of approximately measuring an observable in a given state. A naive version of Heisenberg's origin argument of error and disturbance in measuring jointly two canonical observables was shown to be not universally valid, which has been verified by some recent experiments such as Vienna experiment  \cite{wien}, Toronto experiment \cite{toronto}, and many others \cite{exp13,exp14a,exp14b}. Some refinements of Ozawa's uncertainty relation  \cite{Hall2004, Weston2013,Branciard2013} as well as the tradeoff between different error types \cite{lu} have also been proposed. Especially, Branciard \cite{Branciard2013} proposed a tight error tradeoff in the sense that those errors saturating the bound are realizable, which has been experimentally tested recently \cite{exp14a}.

Werner \cite{werner} and Busch, Lahti, and Werner (BLW) \cite{blw1,blw2} proposed a different kind of errors, namely the worst-case error, to characterize the intrinsic precision of the measuring apparatus since it is independent of the input states. The original Heisenberg's uncertainty relation is reestablished in a product form for continuous variables \cite {blw1} and in an additive form for a qubit \cite{blw2}. However the exact extent to which how well we can measure two incompatible observables has not yet been found even for the simplest case, two qubit observables.  In this Letter we shall find the exact tradeoff between two worst-case errors in measuring jointly two observables of a qubit. To attain those optimal errors we also construct explicitly the optimal joint measurements, which can be readily implemented without introducing any ancilla nor any entanglement. 

Consider two observables $A=\vec a\cdot\vec\sigma$ and $B=\vec b\cdot\vec\sigma$ of a qubit with unit Bloch vectors $|\vec a|=|\vec b|=1$, where $\vec\sigma$ denotes the vector formed by three Pauli matrices. Two noncommuting observables cannot be jointly measured but they can be measured in a single apparatus by allowing errors in measuring them.
Let $\{M_{\mu\nu}\}$ with  ${\mu,\nu}=\pm1$ be an arbitrary positive-operator valued measure (POVM)  with four outcomes, i.e., $M_{\mu\nu}\ge0$ and $\sum_{\mu\nu}M_{\mu\nu}=I$.
Its two marginal POVMs
\begin{eqnarray*}\sum_\nu M_{\mu\nu}&=&\frac{I+\mu(x+\vec m\cdot\vec\sigma)}2:=O_\mu(x,\vec m)\\
\sum_\mu M_{\mu\nu}&=&\frac{I+\nu(y+\vec n\cdot\vec\sigma)}2:=O_\nu(y,\vec n)
\end{eqnarray*}
define two unsharp observables that are jointly measurable, e.g., by the POVM $\{M_{\mu\nu}\}$. Here $x,y$ are two real numbers, referred to as biasedness, and $\vec m$, $\vec n$ are two Bloch vectors satisfying $|\vec m|+|x|\le 1$ and $|\vec n|+|y|\le 1$. For two most general unsharp observables as given above the necessary and sufficient condition for the joint measurability is given in \cite{stano, busch08,jm}. If we regard these two marginal POVMs as unsharp measurements of two sharp observables $A$ and $B$, then these two sharp observables are approximately measured in a single apparatus. The main issue becomes how to quantify the error of measuring an ideal observable by an unsharp observable.

Werner \cite{werner} and also BLW \cite{blw1,blw2} introduced a kind of worst-case error that quantities the largest distance (over all possible input states) between two probability distributions obtained by sharp and unsharp measurements of the given observable. In the case of a qubit the worst-case error in measuring observable $A$ or $B$ by unsharp measurements $\{O_\mu(x,\vec m)\}$ or $\{O_\nu(y,\vec n)\}$ reads
\begin{equation*}
\epsilon_a=|x|+|\vec a-\vec m|\quad \mbox{or}\quad \epsilon_b=|y|+|\vec b-\vec n|,
\end{equation*} respectively. For the sake of convenience the worst-case error given here is half of the error introduced in \cite{blw2}. Comparing with other definitions of errors, e.g., root-mean-square errors, the worst-case errors are state-independent and thus are capable of quantifying how precise a measuring apparatus is in the worst case. A tight lower bound of the error sum is obtained recently \cite{blw2} and our main result is a complete characterization of all the possible and forbidden values of these two errors.

{\it Main results.--- }All the allowed errors $(\epsilon_a,\epsilon_b)$ in measuring jointly two observables $A=\vec a\cdot\vec\sigma$ and $B=\vec b\cdot\vec\sigma$ of a qubit, with $|\vec a\times\vec b|=\sin\theta$, form a convex set shown as the region with red boundaries in Fig.1. This convex set is completely characterized by the upper bound $\epsilon_{a},\epsilon_{b}\le 2$ and a family of lower bounds 
\begin{eqnarray}\label{e}
{\epsilon_a\sin \varphi+\epsilon_b\cos\varphi}&\ge&\sqrt{1+\sin\theta\sin2\varphi}-1,%\frac{\sqrt2z(1+\sin\theta)}{\sqrt{1+\sin\theta(2z^2-1)}}-\sqrt2z:=L(z)\nonumber \\
%&\ge&L(1)=\sqrt{2(1+\sin\theta)}-\sqrt2
\end{eqnarray}
which is plotted as a straight line in Fig.1 for an arbitrary
$\varphi\in[0,\pi/2]$. BLW's bound for the error sum is reproduced as a special case  $\varphi=\pi/4$, which is obviously not attainable in the case of asymmetric errors.  The exact error tradeoff,
shown as the red curve in Fig.1, is the envelop of this family of lower bounds
\begin{subequations}\label{E}%
\begin{eqnarray}
\mE_a=\frac{\sin \varphi+\sin\theta\cos\varphi}{\sqrt{1+\sin \theta\sin 2\varphi}}-\sin \varphi,\\
\mE_b=\frac{\cos \varphi+\sin\theta\sin\varphi}{\sqrt{1+\sin \theta\sin 2\varphi}}-\cos \varphi,
\end{eqnarray}
\end{subequations}
with $\varphi\in[0,\pi/2]$. All those pairs of errors  $(\epsilon_a,\epsilon_b)$ inside the gray-shaded region below the optimal error-tradeoff curve in Fig.1 are forbidden, violating at least one lower bound in Eq.(\ref{e}), whereas all the other errors $\le 2$ are allowed.  The boundary errors $(\mE_a,\mE_b)$ on the optimal error tradeoff curve are attainable (see below) and optimal: on the one hand, if the error $\epsilon_a$ of measuring $A$ is {\it at most} $\mE_a$, i.e., $\epsilon_a\le \mE_a$, then the error of measuring $B$ jointly is {\it at least} $\mE_b$, i.e., $\epsilon_b\ge\mE_b$; on the other hand, if the error of measuring $B$ is is {\it at most} $\mE_b$, i.e., $\epsilon_b\le\mE_b$, then the error of measuring $A$ jointly is {\it at least} $\mE_a$, i.e., $\epsilon_a\ge\mE_a$. 

\begin{figure}
\includegraphics[scale=0.8]{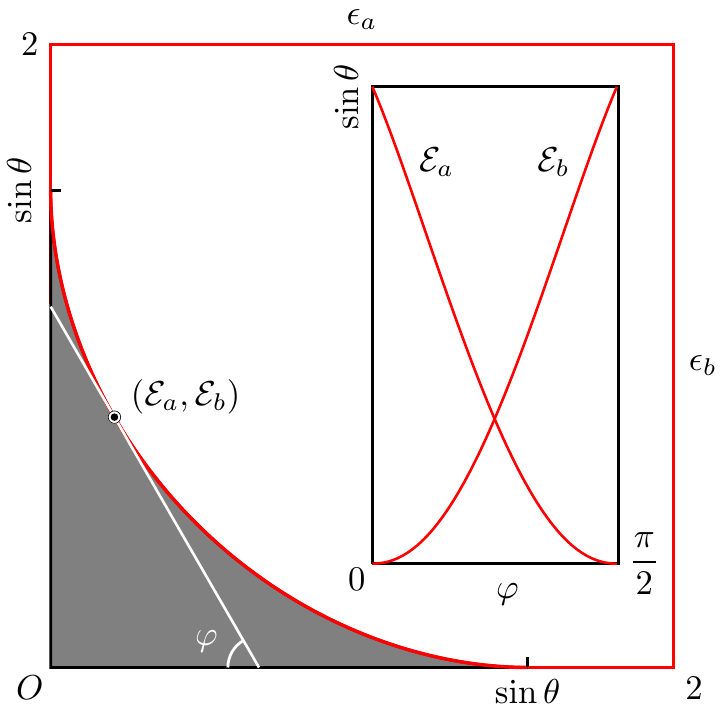}
\caption{(Color online) All the allowed errors $(\epsilon_a,\epsilon_b)$ form a convex set bounded by red curve and straight lines while all the errors in the gray-shaded region are forbidden with boundary given by the curve $(\mE_a,\mE_b)$. Straight line tangent to the curve indicates the lower bound corresponding to $\varphi$.}
\end{figure}

As shown in the inset in Fig.1, both optimal errors $\mE_a$ and $\mE_b$ are  monotonous functions of $\varphi\in [0,\pi/2]$. Thus for any given pair of optimal errors $(\mE_a,\mE_b)$ with $\mE_{a}(\mE_{b})\in[0,\sin\theta]$ there is a unique $\varphi$ determined by Eq.(\ref{E}). The following four operators
\begin{equation}\label{M}
M_{\mu\nu}=\frac{(1+\mu\nu c)I+(\mu\vec m+\nu\vec n)\cdot\vec \sigma}4,
%\, P_{\pm}=\frac{1\pm \sqrt{1-r^2}}2
\end{equation}
with $\mu,\nu=\pm1$ and
\begin{subequations}
\begin{eqnarray}
&\displaystyle c=\frac{\cos\theta}{\sqrt{1+\sin\theta\sin2\varphi}}&\label{c}\\
&\displaystyle\vec m=\frac{\vec a\left(\mE_b+(1-c^2)\cos\varphi\right)\sin\varphi+\vec b\: c\,\mE_a\cos\varphi}{\sin\theta},&\\
&\displaystyle\vec n=\frac{\vec b\left(\mE_a+(1-c^2)\sin\varphi\right)\cos\varphi+\vec a\,c \,\mE_b\sin\varphi}{\sin\theta},&
\end{eqnarray}
\end{subequations}
define a POVM with two marginal POVMs $\{O_\mu(0,\vec m)\}$ and $\{O_\nu(0,\vec n)\}$ as unsharp measurements of observables $A$ and $B$, respectively, having exactly the optimal errors  $\mE_a=|\vec a-\vec m|$ and $\mE_b=|\vec b-\vec n|$. These worst-case errors are attained  at the pure states with Bloch vectors pointing to two orthogonal directions $\vec a-\vec m$ and $\vec b-\vec n$ for observables $A$ and $B$, respectively.

If the worst-case error of measuring one observable is large enough, e.g., $\epsilon_{a}\ge\sin\theta$, then the joint measurement of the other observable, e.g., observable $B$, can be error free, i.e., $\epsilon_b=0$. In this case the optimal measurements are  the projective measurement $\{O_\pm(0,\vec b)\}$ for measuring $B$ without any error $\epsilon_b=0$ and  an unsharp measurement $\{O_\pm(\epsilon_a-\sin\theta,\vec b\cos\theta)\}$ for  measuring observable $A$ with error  $\epsilon_a$. These two observables are commuting and therefore jointly measurable. We note that to attain some non-optimal values of errors, e.g., $\epsilon_a=2$ and $\epsilon_b=0$, biased observables have to be employed since in the case of $\epsilon_b=0$ the maximal error of measuring $A$ jointly using an unbiased observable is $2\cos\theta/2<2$. 

{\it Experimental implementations.--- } In general the implementation of a POVM $\{M_{\mu\nu}\}$ necessitates an ancilla and entanglement between the system and ancilla.
To carry out the optimal joint measurement Eq.(\ref{M}), however, one does not need an ancilla nor any entanglement. This economic implementation is made possible by the following identity 
$$M_{\mu\nu}=\frac{1+\mu\nu c}2\frac{1+\mu L_{\mu\cdot\nu}}2,\quad L_\pm=\frac{(\vec m\pm \vec n)\cdot\vec \sigma}{|\vec m\pm \vec n|}.$$
In fact if we 
 perform two project measurements of sharp observables $L_\pm$ on the given qubit randomly with probabilities $P_\pm=(1\pm c)/2$, then,  in an arbitrary state $\varrho$, the probability $P(s,\mu)$ of obtaining an outcome $\mu=\pm1$ by measuring $L_s$ with $s=\pm1$ is equal to $\tr\varrho M_{\mu,s\cdot \mu}.$
It follows that the probability $\sum_s P(s,\mu)$ of obtaining outcome $\mu$, regardless of which observable is measured, is equal to  the long-run statistics obtained by unsharp measurement $\{O_\mu(\vec m)\}$ while the probability $\sum_\mu P(\pm\mu,\mu)$ of $\mu=\pm s$, i.e., the outcome $\mu$ is the same or different from the label $s$ of the observable we choose to measure,  coincides with the long-run statistics for  $\{O_\nu(\vec n)\}$. 

To measure the worst-case errors the joint measurement specified above should be applied to two different states of the qubit. Let $P(+|A)$ and $P(+|B)$ be the probabilities of obtaining outcome $+$ by measuring observable $A$ and $B$ in a given state $\varrho$, respectively.
Then the optimal worst-case errors in measuring jointly $A$ and $B$ read
\begin{eqnarray*}
\mE_a&=&[P(+|A)-P(+,+)-P(-,+)]_{{\scriptsize\tr}\varrho\vec\sigma\propto \vec a-\vec m},\\
\mE_b&=&[P(+|B)-P(+,+)-P(-,-)]_{{\scriptsize\tr}\varrho\vec\sigma\propto \vec b-\vec n}.
\end{eqnarray*}
 
Various experiments have been carried out to test Heisenberg's uncertainty relations for the joint measurement of two orthogonal observables, i.e., $\theta=\pi/2$, using the root-mean-square errors. These experiments can be readily adapted to test our uncertainty relation. In this case the optimal error-tradeoff curve Eq.(\ref{E}) becomes a quarter of a unit circle given by parametric equation 
\begin{equation}\label{E2}
\mE_a=1-\sin\varphi,\quad \mE_b=1-\cos\varphi,\quad 0\le\varphi\le\frac \pi2.
\end{equation} 
While Vienna experiment \cite{wien} is not optimal, as pointed out in \cite{blw2} even in the case of symmetric errors, Toronto experiment \cite{toronto} does provide an example of optimal joint measurement of two orthogonal qubit observables. In that experiment, observables $\sigma_z$ and $\sigma_x$ are approximated by two jointly measurable unsharp observables
$$\frac{I\pm\sigma_x\sin\varphi }2,\quad \frac{I\pm\sigma_z\cos\varphi }2.$$
It is clear that the worst-case errors are exactly the optimal values Eq.(\ref{E2}). Originally the joint measurement is realized with the help of an ancilla and entanglement. Equivalently it can  be realized by performing two projective measurements along directions $(\sin\varphi,0,\pm\cos\varphi)$ randomly to the system.

{\it Proof.--- }First of all the upper bound, e.g., $\epsilon_a\le 2$ is due to the fact, e.g., $|\vec a-\vec m|\le 1+|\vec m|$. We note that whenever a pair of errors $(\epsilon_a,\epsilon_b)$ is attainable then all pairs of errors $(\epsilon_a^\prime,\epsilon^\prime_b)$ with $2\ge\epsilon_a^\prime\ge\epsilon_a$ and $2\ge\epsilon_b^\prime\ge\epsilon_b$ are attainable. This is because the  measurement $\{I,0\}$, which has the largest error 2, is jointly measurable with all observables. Therefore we have only to find out all those errors $(\mE_a,\mE_b)$ such that for given error $\mE_a$ of measuring $A$ the smallest error in measuring $B$ jointly is $\mE_b$ {\it and} at the same time for given error $\mE_b$ in measuring $B$ the smallest error in measuring $A$ jointly is $\mE_a$. 
 
Then, as noted by BLW \cite{blw2}, the optimal measurements leading to an optimal pair of errors have to be unbiased, i.e., $x=y=0$. This is because if two unsharp observables $\{O_\mu(x,\vec m)\}$ and $\{O_\nu(y,\vec n)\}$ are jointly measurable, then two unsharp observables $\{O_\mu(-x,\vec m)\}$ and $\{O_\nu(-y,\vec n)\}$ can also be measured jointly due to, e.g., the necessary and sufficient condition for the joint measurement of two unsharp observables \cite{jm}. As a result two unbiased observables arising from their convex combinations $O_\mu(0,\vec m)=[O_\mu(x,\vec m)+O_\mu(-x,\vec m)]/2$ and $O_\nu(0,\vec n)=[O_\nu(y,\vec n)+O_\nu(-y,\vec n)]/2$ are also jointly measurable and give rise to smaller errors.

\begin{figure}
\includegraphics[scale=0.7]{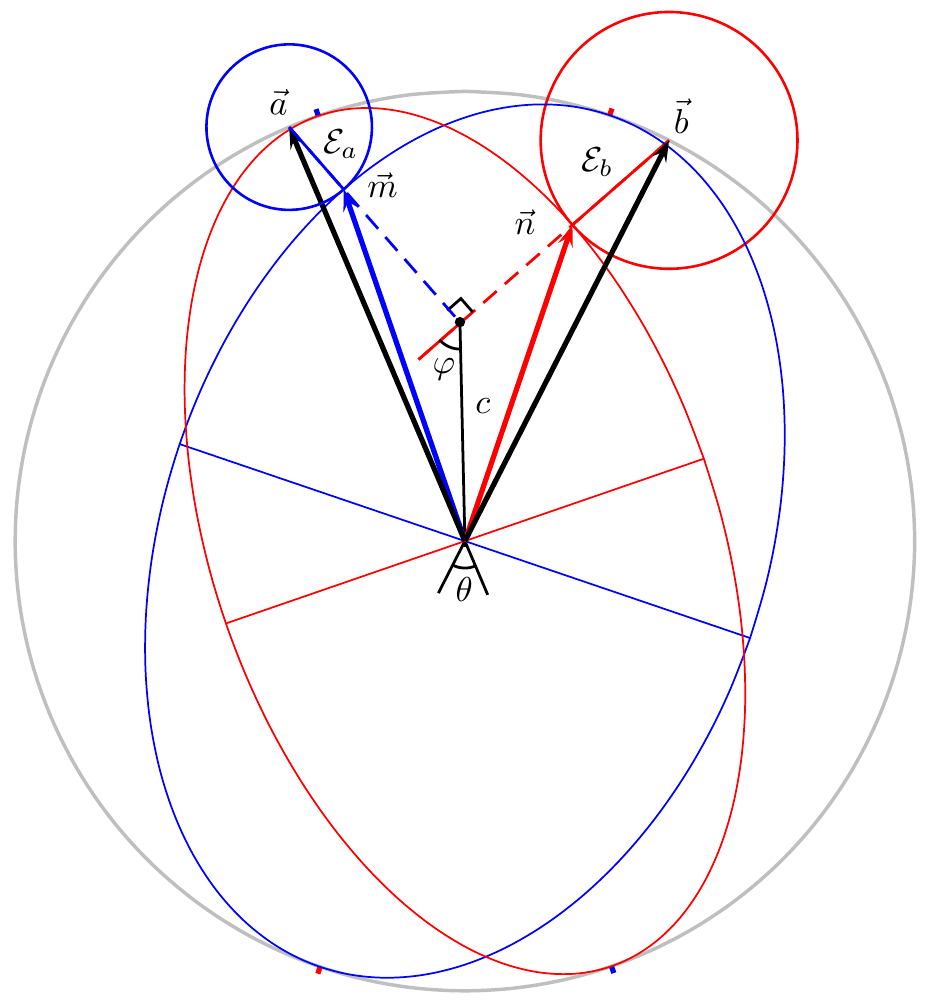}
\caption{(Color online.) The optimal joint measurement has two marginal unbiased observables with Bloch vectors $\vec m$ and $\vec n$. All unbiased observables along directions within the blue (or red)  ellipse, whose foci are located at $\pm \vec n$ (or $\pm \vec m$) with a unit major axis, are jointly measurable with observable along direction $\vec n$ (or $\vec m$). Thus the error is optimal when the blue (or red) circle, with its radius being the optimal error $\mE_a$ (or $\mE_b$), is tangent the to blue (or red) ellipse. }
\end{figure}

 Let two unbiased observables $\{O_\mu(\vec m)\}$  and $\{O_\nu(\vec n)\}$ be the optimal unsharp measurements of observables $A$ and $B$, i.e., attaining smallest possible errors $(\mE_a,\mE_b)$. Two unbiased observables $\{O_\mu(\vec m)\}$  and $\{O_\nu(\vec n)\}$ are jointly measurable if and only if \cite{buschd}
\begin{equation}\label{jmc}
2\ge|\vec m+\vec n|+|\vec m-\vec n|:=h(\vec m,\vec n),
\end{equation}
which defines a region enclosed by an ellipsoid $E_m$ or $E_n$ for given $\vec m$ or $\vec n$, respectively.  The minimal distance from any vector outside the ellipsoid to the enclosed region must be achieved at the boundary. Thus we can assume that $h(\vec m,\vec n)=2$ for optimal joint measurement. Furthermore
to ensure that errors $(\mE_a,\mE_b)$ are optimal,
two spheres $|\vec a-\vec m|=\mE_a$ and $|\vec b-\vec n|=\mE_b$ should be tangent to  ellipsoids $E_n$ and $E_m$ respectively. This amounts to the following two conditions
\begin{subequations}\label{cir}
\begin{eqnarray}
\vec a-\vec m&\propto &\nabla_{\vec m}h(\vec m,\vec n)=\frac{\vec m+\vec n}{|\vec m+\vec n|}+\frac{\vec m-\vec n}{|\vec m-\vec n|},\\
\vec b-\vec n&\propto &\nabla_{\vec n}h(\vec m,\vec n)=\frac{\vec m+\vec n}{|\vec m+\vec n|}+\frac{\vec n-\vec m}{|\vec m-\vec n|},
\end{eqnarray}
\end{subequations}
from which it follows $(\vec a-\vec m)\cdot (\vec b-\vec n)=0$ and that four vectors $\vec a,\vec b,\vec m,\vec n$ are coplanar. Their relations are shown schematically in the plane spanned by $\vec a$ and $\vec b$ in Fig.2, in which we have taken $\theta\in[0,\pi/2]$. For other values of $\theta$, we note that the optimal errors for observables $A,B$, whose optimal measurements are given by $(\vec m,\vec n)$, are identical to those for observables $A$ and $-B$, with optimal measurement being given by $(\vec m,-\vec n)$.  

Let $C=\vec m\cdot\vec n$ and from the joint measurability condition $h(\vec m,\vec n)=2$ it follows $1+C^2={\vec m^2+\vec n^2}$ and $|\vec m\pm\vec n|=1\pm C$. By introducing 
\begin{equation}
\varphi=\arcsin \frac{\sqrt{1-n^2}}{\sqrt{1-C^2}}
\end{equation}
and taking into account the definitions of worst-case errors $|\vec a-\vec m|=\mE_a$ and $|\vec b-\vec n|=\mE_b$,
we obtain
\begin{eqnarray}\label{ab}
\vec a=\vec m+\frac{\mu\mE_a(\vec m-C\,\vec n)}{(1-C^2)\sin\varphi},\quad
\vec b=\vec n+\frac{\nu\mE_b(\vec n-C\,\vec m)}{(1-C^2)\cos\varphi}
\end{eqnarray}
 from Eq.(\ref{cir}) with $\mu,\nu=\pm1$ being arbitrary. From $|\vec a|=1$, $|\vec b|=1$, and $|\vec a\times\vec b|=\sin\theta$  we obtain, respectively, 
\begin{eqnarray*}
\mE_a&=&\textstyle\sqrt{1-C^2\cos^2\varphi}-\mu\sin\varphi,\\
\mE_b&=&\textstyle\sqrt{1-C^2\sin^2\varphi}-\nu\cos\varphi,\\
C^2&=&\frac{\cos^2\theta}{{1+\mu\nu \sin\theta\sin2\varphi}}.
\end{eqnarray*}

If $\mu=\nu=-1$ then both two errors are obviously larger than those in the case of $\mu=\nu=1$. If $\mu=-\nu=1$ then $\sin2\varphi\le \sin\theta$ since $|C|\le 1$, meaning that we have either $\varphi\le\theta/2$, in which case $\mE_b$ is a decreasing function of $\varphi$, or $\varphi\ge(\pi-\theta)/2$, in which case $\mE_b$ has a single critical point determined by $\mE_b^\prime(\varphi)=0$ that is a local maximum. As a result the minimal value of $\mE_b$ is taken at $\varphi=\theta/2$ or $\varphi=(\pi-\theta)/2$ or $ \varphi=\pi/2$, which is equal to $\sin\theta$, i.e., $\mE_b\ge\sin\theta$, while in the case of $-\mu=\nu=1$ we have $\mE_a\ge \sin\theta$. In both cases the other observable can be jointly measured without error. Therefore one should choose  $\mu=\nu=1$ to ensure that $(\mE_a,\mE_b)$ is optimal, which leads to Eq.(\ref{E}). Moreover we have $C=c$ as defined in Eq.(\ref{c}) since $\vec a\cdot\vec b=\cos\theta$. 
The tangents to the optimal error-tradeoff curve Eq.(\ref{E}) immediately give rise to the  lower bounds Eq.(\ref{e}) by noting that
$$\frac{d\mE_a}{d\sin\varphi}=\frac{c^3}{\cos\theta}-1,\quad \frac{d\mE_b}{d\mE_a}=-\tan\varphi.$$
 It also follows that both optimal errors $\mE_a,\mE_b$ are monotonous functions of $\varphi$ and one of the optimal error, e.g., $\mE_b$, is a decreasing convex function of the other, e.g., $\mE_a$. On the other hand the envelop of the family of straight lines in Eq.(\ref{e}) can be shown easily to be the optimal error-tradeoff curve Eq.(\ref{E}).

{\it Conclusions and discussions.--- }Lying at the heart of Heisenberg's uncertainty relation is the question of how well we can measure jointly two incompatible observables. In the case of qubit we provide a complete answer to this question by providing the exact error tradeoff  between the worst-case errors in measuring two observables. To attain each optimal pair of optimal errors we have explicitly constructed a joint measurement that is realizable without the need of an ancilla and any entanglement. Those experimental tests of state-dependent Heisenberg's uncertainty relations for measurements can be readily adapted to test our exact error tradeoff. 

{\it Acknowledgement.--- }This work is funded by the Singapore Ministry of Education (partly through the Academic Research Fund Tier 3 MOE2012-T3-1-009).

\newpage

\end{document}